\documentstyle[12pt,aaspp4,psfig]{article}
\def\kms{{\rm km/s}}
\def\um{\mu {\rm m}}
\def\kpc{{\rm kpc}}
\def\deg{^\circ}
\def\@journalname{Astrophysical Journal}
\def\accepted#1{\gdef\@accptdate{#1}}
\accepted{\relax}
\def\journalid#1#2{\gdef\@jourvol{#1}\gdef\@jourdate{#2}}
\def\articleid#1#2{\gdef\@startpage{#1}\gdef\@finishpage{#2}}
\begin{document}

\title{The Milky Way, The Local Galaxies \& the IR Tully-Fisher Relation}
\author{Sangeeta Malhotra\altaffilmark{1,}\altaffilmark{3}, David N Spergel%
\altaffilmark{2,}\altaffilmark{4}, James E Rhoads\altaffilmark{2}, Jing Li%
\altaffilmark{1}}

\begin{abstract}
Using the near infrared fluxes of local galaxies derived from
Cosmic Background Explorer (COBE)/Diffuse Infrared Background Experiment
(DIRBE)\footnote{COBE data sets were developed by the NASA Goddard
Space Flight Center under the guidance of the COBE Science Working
Group and were provided by the NSSDC.}  J(1.25 $\um$) K (2.2 $\um$) \&
L (3.5 $\um$) band maps and published Cepheid distances, we construct
Tully-Fisher diagrams for the nearby galaxies.  The measured dispersions
in these luminosity-linewidth diagrams are remarkably small: $\sigma_J
= 0.09$ magnitudes, $\sigma_K = 0.13$ magnitudes, and $\sigma_L =
0.20$ magnitudes.  These dispersions include contributions from both
the intrinsic Tully-Fisher relation scatter and the errors in
estimated galaxy distances, fluxes, inclination angles, extinction
corrections, and circular speeds.  For the J and K bands, Monte Carlo
simulations give a 95\% confidence interval upper limit on the true
scatter in the Tully-Fisher diagram of $\sigma_J \le 0.35$ and $\sigma_K
\le 0.45$.

We determine Milky Way's luminosity and place it in the
Tully-Fisher diagram by fitting a bar plus exponential disk model of
the Milky Way to the all-sky DIRBE maps.  For ``standard'' values of
its size and circular speed (Sun-Galactic center distance $R_0 = 8.5
\kpc$ and $\Theta_0 =220 \kms$), the Milky Way lies within 
$1.5 \sigma$ of the TF relations.

We can use the TF relation and the Cepheid distances to nearby bright
galaxies to constrain $R_0$ and $\Theta_0$: $-\log\left(R_0 / 8.5
\kpc\right) +1.63\log\left(\Theta_0 / 220 \kms\right) = 0.08 \pm 0.03$.
Alternatively, we can fix the parameters of the Galaxy to their
standard values, ignore the Cepheid zero-point, and use the
Tully-Fisher relation to determine the Hubble Constant directly: $H_0
= 66 \pm 12$ km/s/Mpc.

We have also tested the Tully-Fisher relation at longer wavelengths,
where the emission is dominated by dust.  We find no evidence for a
Tully Fisher relation at wavelengths beyond 10$\mu$m.  The tight correlation
seen in L band suggests that stellar emission dominates over the 3.3
$\mu$m PAH emission.

\keywords{Galaxy: fundamental parameters
Galaxy: general
galaxies: distances and redshifts}
\end{abstract}

\altaffiltext{1}{IPAC, 100-22, California Institute of Technology, Pasadena,
CA 91125} \altaffiltext{2}{Princeton University Observatory, Princeton NJ
08540} \altaffiltext{3}{NRC Fellow at NASA Jet Propulsion Laboratory}
\altaffiltext{4}{Department of
Astronomy, University of Maryland, College Park, MD 20742} 

\section{Introduction}

The Tully-Fisher (TF) relation between the luminosity and linewidth of
spiral galaxies (Tully \& Fisher 1977) has been used extensively as a
distance indicator and to map large scale flows of galaxies (cf., Strauss \&
Willick 1995 and Jacoby et al. 1992 for reviews). Its usefulness as a
distance indicator is limited by the intrinsic scatter of the relation. This
scatter is lowest in redder bands
where dust extinction is low:  H band ($1.65 \um$) (Aaronson, Huchra
\& Mould 1979, Aaronson, Mould \& Huchra 1980, Aaronson et al. 1989, 
Freedman 1990, Pierce \& Tully 1992) and I band ($0.90 \um$)
(Bernstein et al. 1994).
In this paper, we extend the TF relation to longer wavelengths. The
DIRBE experiment, with its excellent calibration and large beam width,
is ideal for measuring the total flux of galaxies in the local galaxies.
We describe our galaxies data set and present the results of our
analysis in section 2.

The Milky Way has often been deemed unsuitable for zero point
calibration of the TF relationship, mainly because of difficulties in
estimating its total luminosity.  Such difficulties can be overcome at
infrared wavelengths, where dust absorption is small.  In section 3,
we use a three-dimensional model of the Milky Way based on the DIRBE
J, K, and L band maps to obtain a measurement of the Galaxy's
luminosity.  This luminosity can be used to place the Milky Way on the
TF diagram with nearby galaxies, to constrain Galactic parameters, and
to obtain an independent calibration of the Hubble constant.

\section{Nearby Galaxies}

For this study, we chose a sample of nearby bright spiral galaxies with
sizes not too much smaller than the beam, measured Cepheid distances, no
bright stars in a $1^\circ$ field centered on the galaxy,
and with inclination greater than 45$\deg$.
We also required that the measured DIRBE flux from the galaxy exceed the
week-to-week fluctuations in the DIRBE data. The galaxies used are M31, M33,
M81, NGC 300, and NGC 2403. A sixth galaxy, NGC 247 shows a marginal
detection and meets our selection criterion in only 2 bands (J and L). We
will do our analysis both with and without NGC 247. Barring M31 and M33,
these galaxies are smaller than the  $0.7\deg \times 0.7\deg$  DIRBE beam.
M33 is slightly larger than the beam, and M31 and Milky Way are
extended objects compared to the beam.

The measured fluxes were corrected for extinction using the
prescription of Tully \& Fouqu\'e (1985), applied to our sample using
inclination angles from Pierce \& Tully (1992) and the standard
extinction law from Mathis (1990).  These corrections were never large,
attaining maximum values of 0.18, 0.07, and 0.03 magnitudes in J, K,
and L for M31. The distances and distance errors were taken from
published literature as follows: M31 (Freedman 1990), M33 (Freedman,
Wilson \& Madore 1991), M81 (Freedman et al. 1994, Hughes et al.
1994), NGC247 (Catanzarite, Freedman \& Madore 1996), NGC300 (Freedman
1990) and NGC2403 (Freedman 1990). Line widths for these galaxies are
taken from Pierce \& Tully (1992) and Freedman (1990).

\subsection{Flux Extraction}

DIRBE was one of the three experiments on board the COBE mission.
This experiment observed the entire sky in ten wavebands (from 1.25
$\mu$m to 240 $\mu$m) over a nine month period.  Each point on the sky
was observed many times, 
at a range of solar elongation angles. 
This gives a variable viewing geometry from within the interplanetary
dust cloud, which is a major source of foreground emission at these
wavelengths. A detailed description of the COBE mission can be found
in Boggess et al. (1992). Details of DIRBE data processing,
calibrations and photometry are given in the DIRBE Explanatory
Supplement (Hauser et al. 1995).

The fluxes of the galaxies were derived from the weekly sky maps
provided by the DIRBE team. Intensity in each pixel of the weekly map,
which is typically half a beam across, is a robust average of all
observations in a week pointing at the region of the sky in that
pixel. Thus the center of the beam for any observation may be up to
half a pixel away from the center of the pixel, and does not in
general lie on the center of the galaxy which lies in that
pixel. Since the beam shape is not flat this can influence the flux
measured for that galaxy.  This error is estimated to be a few to nine
percent in the DIRBE Explanatory Supplement (section 5.6.6).

For each weekly map we fit a first order polynomial to the
sky brightness in a small annulus around the pixel that contains
the galaxy. The sky background was then subtracted from the maps. A point
source can influence the flux levels of neighboring pixels due to the way
the weekly maps were made, so pixels adjacent to galaxies were excluded from
the sky estimate. All the weekly maps which had observations of that part of
the sky were then averaged (after rejecting outliers) to give the the
average flux of the galaxy. The week to week variation of the flux in a
given pixel gives an estimate of the noise in the maps. The flux per pixel
thus obtained in MJy/Sr is multiplied by the beam size (given in the DIRBE
Explanatory Supplement) to derive the flux of the sources.

The flux in a given pixel has contributions from the galaxy,
unresolved Galactic stars and zodiacal light. The zodiacal light
varies from week to week as one sees the same part of the sky through
a different path length of the zodiacal dust cloud near the
Earth. This makes it necessary to remove the background separately for
each week. We assume that the starlight is smoothly varying in space
and can be subtracted as part of the background.  This is true except
in the rare case of a very bright star close to the galaxy.  We
examined the Palomar Observatory Sky Survey plates for such bright
stars near candidate galaxies, and rejected NGC 925 and NGC 4571 from
our sample because of nearby bright stars.  The week-to-week variation
in the flux of the source gives the estimate of errors due to
imperfect zodiacal light subtraction, and also due to the beam being
centered at different positions with respect to the galaxy. We have
used the week-to-week flux variation as a conservative estimate of the
errors in the flux measurement.

M33 is slightly bigger than the beam size, and the single pixel method
described above underestimates its flux. We simulated observations of M33
assuming an exponential disk with a scale length of $6 \arcmin$ 
(Regan \& Vogel 1994) and the beam profile of the DIRBE beam. In these
simulations, we placed the beam randomly within the pixel closest to the
position of M33 center. The same experiment was repeated with a point
source. The flux measured for the M33 simulation was compared to the flux
measured for a point source and a correction factor of -0.55 magnitude was
added to the M33 magnitude. M31 appears extended in the DIRBE maps, so its
flux was measured by summing all the pixels containing the galaxy and
normalizing this sum by the ratio of the pixel area to the beam area.

\subsection{Maximum Likelihood Analysis}

We determined the TF parameters and errors using a maximum likelihood
analysis. This approach is preferable to a $\chi^2$ fit to the data as the
maximum likelihood analysis includes the selection bias due to our magnitude
limit (Willick 1994). In a magnitude limited sample, selection bias implies
that galaxies near the luminosity cut-off are systematically brighter
(Willick 1994). NGC 247, which falls just below our cut-off, behaves just as
expected given this effect: it is brighter than the best fit TF relation. We
perform our analysis both with and without NGC 247.

The maximum likelihood function for our sample is, 
\begin{equation}
P(\vec m|a,b,\sigma_{TF})=\prod_i{\frac{\exp \left\{ -{\frac{\left[
m_i - D_i -(a\eta _i+b)\right] ^2}{2(\sigma _{TF}^2)}}\right\} }{\int_{-\infty
}^{m_{l,i}}\exp \left\{ -{\frac{\left[ m^{\prime }- D_i -(a\eta _i+b)\right] ^2
}{2(\sigma _{TF}^2)}}\right\} dm^{\prime }}}~~.
\end{equation}
Here the observed quantities are the galaxies' magnitudes $m_i$,
Cepheid-based distance moduli $D_i$,
the logarithms of their line widths $\eta _i$, and the magnitude limits 
$m_{l,i}$ (i.e., the amplitude of the week-to-week fluctuations measured at
the galaxy location). The model parameters $a$, $b$, and $\sigma _{TF}$ are
the slope, zero-point, and [total] scatter about the Tully-Fisher
relation.  Note that $\sigma _{TF}$ includes contributions from both
the intrinsic Tully-Fisher relation scatter and the errors in
estimated galaxy distances, fluxes, inclination angles, extinction
corrections, and circular speeds.  The product is evaluated over all
galaxies in the sample, i.e., all galaxies for which $ m_i<m_{l,i}$.


The maximum likelihood fits to the J, K and L band Tully Fisher relations
(Figures 1, 2 \& 3) are: 
\[
J^c=-8.13_{-0.42}^{+0.41}\log (\Delta V/400{\rm km/s})-22.00_{-0.12}^{+0.12}
\]
\[
K^c=-8.59_{-0.65}^{+0.67}\log (\Delta V/400{\rm km/s})-23.01_{-0.11}^{+0.11}
\]
\[
L^c=-9.01_{-0.94}^{+0.96}\log (\Delta V/400{\rm km/s})-22.99_{-0.17}^{+0.14}
\]
where $J^c,K^c$ and $L^c$ are dust-corrected magnitudes. The
linewidth-luminosity, or Tully Fisher, relation in these bands is very
tight, showing a scatter of $\sigma _{TF}(J)=0.09$ magnitudes in J
band, $\sigma _{TF}(K)=0.13$ magnitudes in K band and $\sigma
_{TF}(L)=0.20$ magnitudes in L band. $\chi ^2$ fits yield smaller (but
less believable) values for $\sigma $. The minimization of the
likelihood function was done with the Numerical Recipes (Press et
al. 1992) AMOEBA program. Investigation of the likelihood surface
showed that it was well-behaved near its minimum.  A small scatter in
the TF relation of 5 galaxies implies a small intrinsic scatter in the
TF relation for all galaxies. We confirm this for our sample with a
Monte-Carlo simulation with 1000 realizations of the local galaxies
sample. We generated 5 galaxies with the reported $\eta _i$, and our
best fit slopes.  We varied the value of $\sigma _{TF}$ in our
simulations. For a true $\sigma _{TF}(J)=0.35$, we recovered a scatter
as small as observed in less than 5\% of the simulations. For $\sigma
_{TF}(J)=0.55$, we recovered such a small scatter in only 1\% of the
simulations.  The quoted errors on the TF slope and zero-point are the
95\% confidence intervals from the Monte-Carlo simulations.

If we include NGC 247 in the sample, then the maximum likelihood fits are 
\[
J^c=-7.78\log(\Delta V/400 {\rm km/s})-22.07 ~~,
\]
\[
K^c=-8.90\log(\Delta V/400 {\rm km/s})-22.99 ~~,
\]
\[
L^c=-8.76\log(\Delta V/400 {\rm km/s})-23.00 ~~,
\]
and the corresponding Tully Fisher relation scatter is $\sigma_J=0.15$,
$\sigma_K=0.14$, and $\sigma_L=0.21$ magnitudes. For a true $\sigma_{TF}(K) =
0.45$, we found such small scatter in less than 5\% of simulations. For
$\sigma_{TF}(K) = 0.7$, we found such smaller scatter in only 1\% of
simulations. 

The dust emission at longer wavelengths 60-240 $\mu {\rm m}$ does not
show a luminosity-linewidth relation. From this and from the tight L
band Tully Fisher correlation, we conclude that the non-stellar or
dust contribution to the L band is not large. This is consistent with
recent estimates of 8-16 \% contribution of the 3.3 $\mu {\rm m}$ PAH
feature to the L band (Bernard et al. 1994)

While giant stars are thought to contribute most of the light in near IR
bands, supergiants can make an important contribution in the spiral arms
(Rhoads 1996). This contribution does not appear to be a source of
significant scatter in the TF relation.

\section{The Milky Way}

We cannot directly use the total flux from the Milky Way (MW) as
regions at different distances contribute to that flux. We estimated
the luminosity density of the MW from the best fitting model to the
DIRBE data. The MW was modelled as a sum of an exponential disk and a
bar (Spergel, Malhotra \& Blitz 1996, Paper I). Extinction corrections
were based on a 3-dimensional dust model. Varying modelling parameters
and extinctions changes the MW flux by less than 10
a 10
circular speed $\Theta _0=220\pm 10$ km/s and a distance to the
Galactic center $R_0=8.5\pm 0.5$ kpc (Gunn, Knapp \& Tremaine
1979). Using these values, we obtain absolute magnitudes $J=-23.05$,
$K=-24.06$, and $L=-23.88$. More recent determinations of the
Sun-Galactic center distance using water masers suggest a smaller
distance: $7.1\pm 1.5$ kpc (Reid et al. 1988) and $8.1\pm 1.1$ kpc
(Gwinn, Moran \& Reid 1992). Combining our uncertainties and assuming
a TF slope of $9$ implies an intrinsic uncertainty in the Galaxy's
position in the TF diagram of $0.23$ magnitudes.

If we include Milky Way in the sample, then the maximum likelihood fits are: 
\[
J^c=-8.09\log(\Delta V/400 {\rm km/s})-22.15
\]
\[
K^c=-9.23\log(\Delta V/400 {\rm km/s})-23.03
\]
\[
L^c=-8.91\log(\Delta V/400 {\rm km/s})-23.04
\]
For uniformity, the velocity width of the MW was scaled from the
observed velocity width of M31: $\Delta V (MW)=(220/250) \times \Delta
V (M31)$. The Tully Fisher relation in these bands remains very tight
with $\sigma_J=0.18$, $\sigma_K=0.16$, and $\sigma_L=0.21$. We
repeated the Monte Carlo simulations as described in Section~2.  For a
true $\sigma_{TF}(K) = 0.45$, we found a scatter as small as observed
in less than 5\% of simulations.  For $\sigma_{TF}(K) = 0.57$, we found
such a small scatter in only 1\% of simulations.  With the standard
parameters, the Milky Way is $\sim 1.5 \sigma$ too luminous in J,
$\sim 1 \sigma$ too luminous in  K, 
and $\sim 0.3 \sigma$ too luminous in L. This basic agreement shows
that the Cepheid distance scale is consistent at the 10\% level with
Gunn et al. (1979) distance determination to the Galactic center.

Alternatively, we can use the TF relation obtained for the 5 galaxy
sample, together with our model for the Galactic luminosity to
constrain Galactic parameters:
\[
-5\log\left({\frac{R_0 }{8.5 {\rm kpc}}}\right) +8.13\log\left({\frac{%
\Theta_0 }{220 {\rm km/s}}} \right) = 0.42 \pm 0.16
\]
\[
-5\log\left({\frac{R_0 }{8.5 {\rm kpc}}}\right) +8.59\log\left({\frac{%
\Theta_0 }{220 {\rm km/s}}} \right) = 0.39 \pm 0.19
\]
\[
-5\log\left({\frac{R_0 }{8.5 {\rm kpc}}}\right) +9.01\log\left({\frac{%
\Theta_0 }{220 {\rm km/s}}} \right) = 0.20 \pm 0.24
\]
where the 3 equations correspond to fitting the J, K and L band
relations respectively. The uncertainties include the width of the TF
relation, the quoted uncertainties in the Galactic model (10\%) and a
conservative estimate of systematic errors in point source flux
measurements (9\%). This comparison of the Galaxy with the nearby
galaxies sample suggests that its rotation speed is unlikely to be as
small as 200 km/s.

Finally, we can also use the Galaxy (if we are willing to assume values for
$R_0$ and $\Theta_0$) to obtain a value of the Hubble parameter that is
independent of the zero-point of the Cepheid distance scale (Wright 1994).
If we use the Galaxy as the zero-point of the K band Tully Fisher relation,
we find that 
\[
K^c=-8.59[\log(\Delta V)-2.5]-22.52 \pm 0.29 -5\log\left({\frac{R_0 }{8.5 
{\rm kpc}}}\right) +8.59\log\left({\frac{\Theta_0 }{220 {\rm km/s}}} \right)
\]
where the zero-point error is the sum in quadrature of the uncertainties in
Galactic parameters, the systematic error in the point source flux
measurements,
the uncertainties in the Galaxy model, and the deviation of a typical galaxy
from the TF relation.

In order to extrapolate to $H$ band, where there are measured fluxes
for large numbers of external galaxies (Aaronson et al. 1986), we use
measured H band fluxes for our sample (Freedman 1990, Pierce \& Tully
1990) to determine the dust-corrected $H^c-K^c$ and $J^c-K^c$
color-color relation for the galaxies in this sample. A $\chi^2$ fit
to the data yields $(H^c-K^c) = 3.27 - 2.50 \times (J^c-K^c)$ with a
scatter of 0.18 magnitudes.  This suggests $(H^c-K^c) = 0.75 \pm 0.18$
for the Milky Way and therefore, $H^c = -21.77 \pm 0.34$ for a galaxy
with $\log(\Delta v) = 2.5$. If we use this estimate to set a
zero-point for the H band TF relation and combine it with the Aaronson
et al. (1986) TF study of the local velocity field (their Table 6 and
Equation 4), this yields $H_0 = 66 \pm 12$ km/s/Mpc. This value is
lower than that obtained by using Cepheid distances and the Local
Group to normalize the TF relation (Aaronson et al. 1986, Freedman
1990) as the Galaxy is somewhat over-luminous relative to the rest of
the nearby galaxies sample (for the standard Galactic parameters and
the standard Cepheid distances).  If we assume that $R_0 = 7.1$ kpc
(the smaller maser-based measurement [Reid et al. 1988]) and hold
$\Theta_0$ constant, then the Hubble constant estimate increases to
$79$ km/s/Mpc. If we assume that $R_0 = 7.1$ kpc and hold the angular
rotation rate, $\Theta_0/R_0$ constant, then the Hubble constant
estimate decreases to 58 km/s/Mpc.  Besides the uncertainty in
Galactic parameters, the dominant source of error in these estimates
is the extrapolation from J and K band fluxes to H band.  Thus, a K
band survey of spirals in the Coma cluster would yield a more
definitive value of the Hubble constant.

\section{Conclusions}

We find that TF relation extends to longer wavelengths (2.2 $\mu {\rm m}$
and 3.5 $\mu {\rm m}$) than previously explored. The extinction corrections
(cf. Mathis 1990) at 2.2 $\mu {\rm m}$ and $3.5\mu {\rm m}$ are about half
and one-third as large as for the H-band at $1.65\mu {\rm m}$, so these
bands may be usefully exploited for distance estimation. With the advent of
imaging IR instruments and two major near-IR sky surveys (2MASS and DENIS),
there is also potential to estimate the distances and hence the peculiar
velocity flows for many more galaxies ( $\sim 10^6$) in a greater part of
the sky and nearer to the plane of the Milky Way.

Using Cepheid distances for galaxies in the Local Group (not including
the Milky Way) and assuming standard Galactic parameters, we find that
the Galaxy obeys the TF relation for nearby galaxies with Cepheid
distances. This consistency is an independent check of the distance
scale.

\section{Acknowledgements}

We thank Leo Blitz, George Helou, Jill Knapp, Barry Madore, Nancy Silbermann
and Ned Wright for discussions and suggestions, and David Leisawitz for
answering questions about the DIRBE data. We also thank Michael Strauss for
helpful comments on an earlier version of this manuscript. This work made
use of NASA Extragalactic Database NED. This work was partially supported by
NASA ADP NAG5-269.

\newpage 
\figcaption{
The J band ($1.25 \mu {\rm m}$) Tully-Fisher diagram.  The error bars
for the local galaxies M31, M81, NGC 2403, M33, NGC 300, and NGC 247
represent the $1 \sigma$ uncertainties in the dust-corrected absolute
$J$ band magnitude ($J^c$) due to photometric errors and reported
uncertainties in Cepheid distances, added in quadrature.  The Milky
Way (MW) absolute magnitude assumes a Galactocenteric distance of $8.5
\pm 0.5$ kpc and circular speed $\Theta_0 = 220 \pm 10 $ km/s; its
error bar represents the 0.23 mag uncertainty derived in section~3.
The solid line is the fit to the 5 galaxy sample, excluding NGC 247
(which falls just below our magnitude limit) and the Milky Way.
The dashed line is the fit to the full 7 galaxy sample.
}
\figcaption{As figure~1, but for K band ($2.2 \mu {\rm m}$). }
\figcaption{As figure~1, but for L band ($3.5 \mu {\rm m}$). }

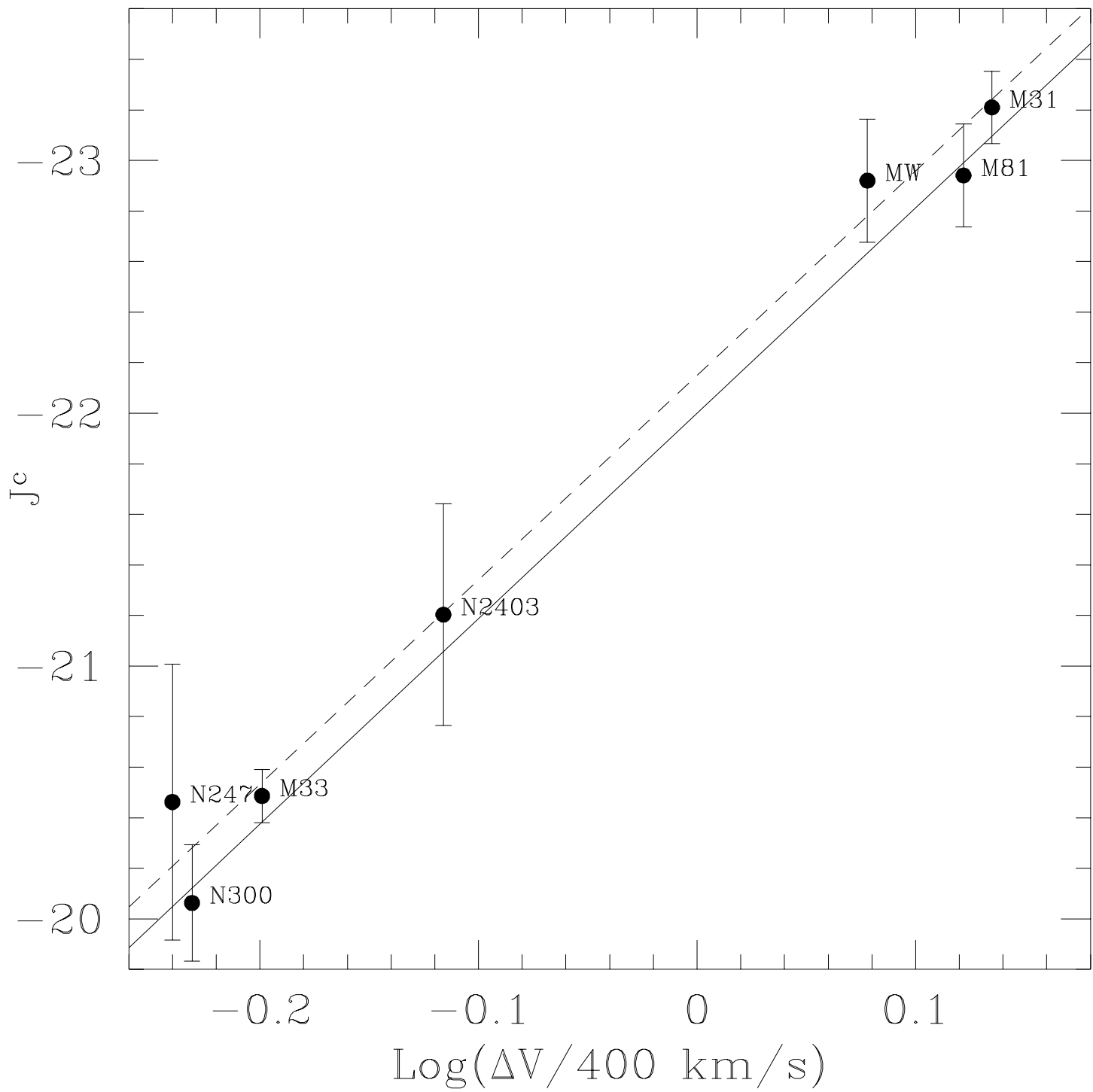
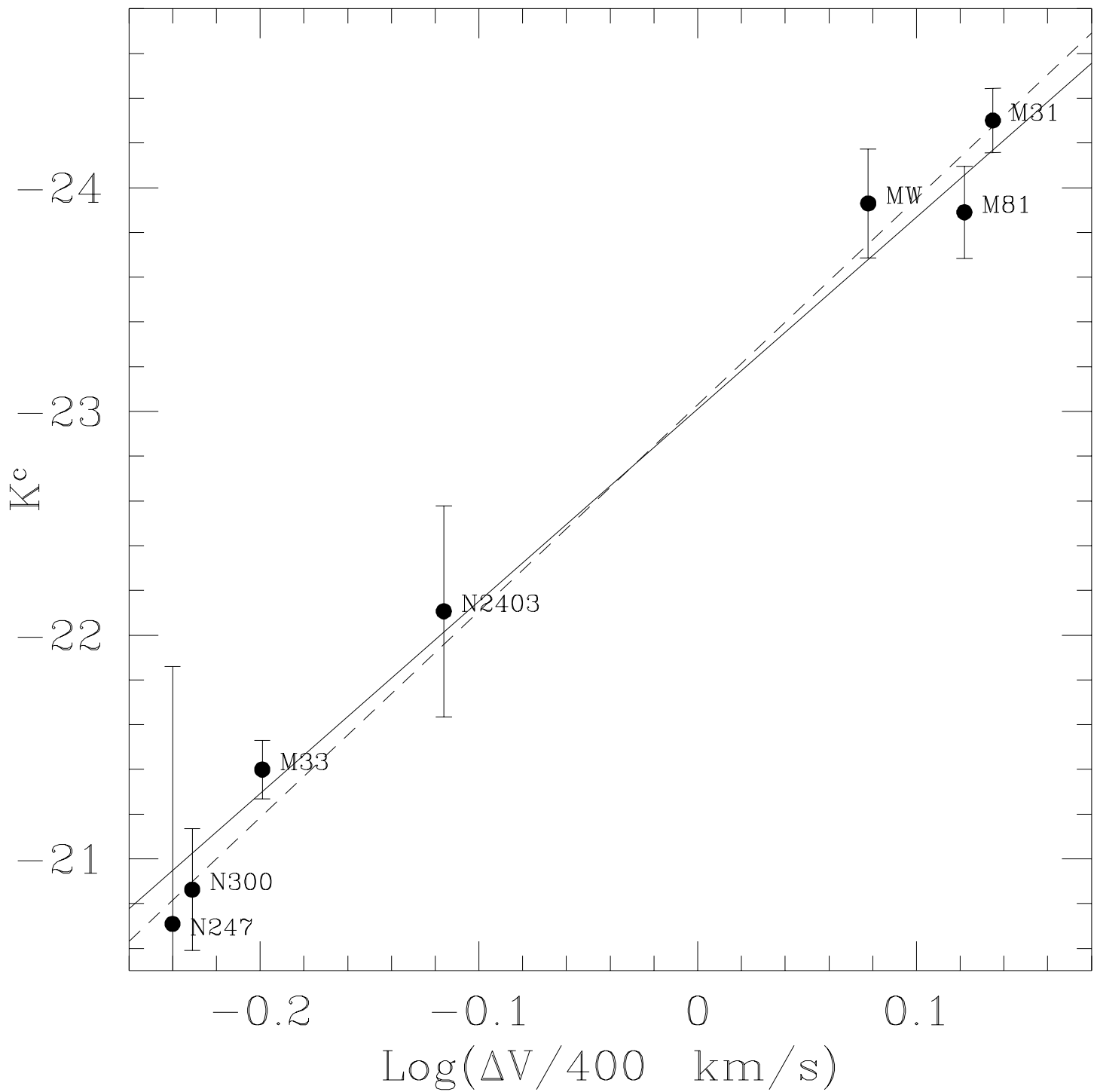
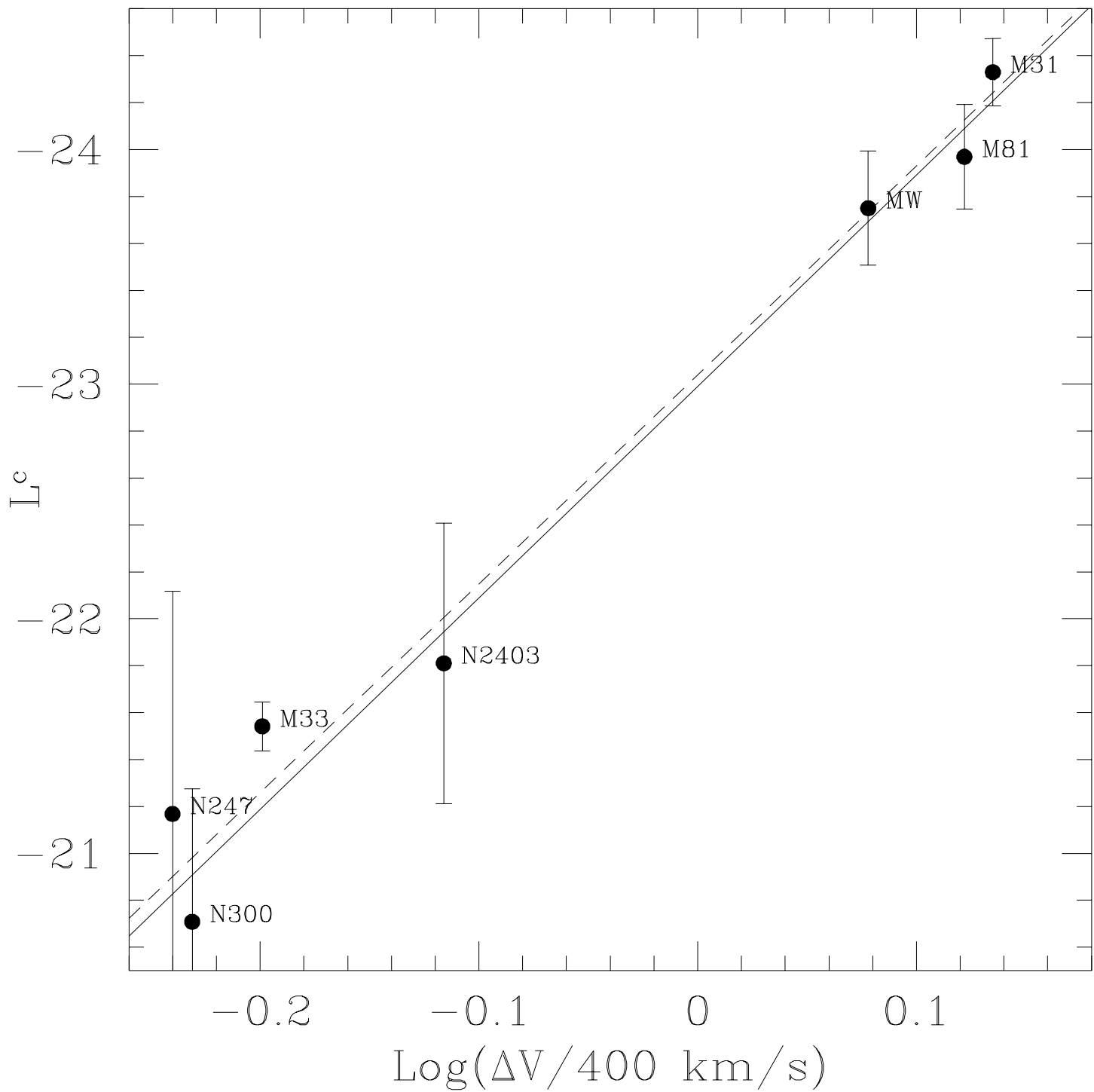
\end{document}